# Traffic Flow Modeling for UAV-Enabled Wireless Networks

Abderrahmane Abada*, Bin Yang*, Tarik Taleb*†
*School of Electrical Engineering, Aalto University, Finland
†Centre for Wireless Communications, University of Oulu, Finland
Email:{abderrahmane.abada, bin.1.yang, tarik.taleb}@aalto.fi

***Abstract*—This paper investigates traffic flow modeling issue in multi-services oriented unmanned aerial vehicle (UAV)-enabled wireless networks, which is critical for supporting future various applications of such networks. We propose a general traffic flow model for multi-services oriented UAV-enable wireless networks. Under this model, we first classify the network services into three subsets: telemetry, Internet of Things (IoT), and streaming data. Based on the Pareto distribution, we then partition all UAVs into three subgroups with different network usage. We further determine the number of packets for different network services and total data size according to the packet arrival rate for the nine segments, each of which represents one map relationship between a subset of services and a subgroup of UAVs. Simulation results are provided to illustrate that the number of packets and the data size predicted by our traffic model can well match with these under a real scenario.**

## I. Introduction

Unmanned aerial vehicle (UAV)-enabled wireless networks are envisioned to be a key component in 5G and beyond wireless networks due to the promising features of UAVs like flexible mobility, large area coverage, and fast deployment [1], [2], [3], [4]. UAVs equipped with various wireless equipments have huge potential to provide numerous services in military and civil fields such as search and rescue operations, firefighting, agriculture, mapping, surveying and Internet of Things (IoT) [5], [6], [7], [8]. However, the dramatically increasing service requirements also poses enormous challenges for future UAV-enabled wireless networks. The latency and the packet loss caused by the poor network design can negatively impact the control of UAVs [9] while the massive data generated by the UAVs can decrease the QoS [10], [11] and the UAVs' communication over the cellular network negatively affect the QoE [12]. To design and deploy such networks, some work has studied the deployment of UAVs' service across the network [13], [14], [15] and using optimized service placement [16] but it stills of a great importance to model the UAV traffic flows for different services.

The existing works on the study of traffic flow mainly focus on the non-self-similar models [17], [18], [19] and self-similar models [20], [21], [22]. The non-self-similar models consist of Poisson, compound Poisson and Markov-modulated Poisson models. The authors in [17] investigated the Poisson traffic model. Under the model, the traffic flow are regarded as a Poisson process, whereby the inter-times of packet arrival are exponentially distributed with a fixed rate parameter. The work in [18] employed the compound Possion traffic model to characterize the batches of packets delivery [18]. This model considers the batch parameter besides the rate parameter. The Markov-modulated Poisson traffic model was also used to depict the time-varying packet arrival rate and capture some of the important correlations between the inter-times of packet arrival while still remaining analytically tractable [19]. On the other hand, the self-similar traffic models consist of Fractional Brownian Motion, Chaotic maps, and Pareto distributed traffic models. The Fractional Brownian Motion was used to model the traffic models in [20], whereby the traffic flow is a continuous-time Gaussian process. The work in [21] illustrated that the Chaotic maps model seen as continuous-state Markov process can characterize the traffic flow for various network services. The authors in [22] investigated the Pareto distribution process based traffic model. The Pareto distribution process can produce independent and identically distributed inter-times of packet arrival.

Notice that the non-self-similar models consider an ideal scenario without bursty traffic. The bursty traffic usually occurs in these communication scenarios of voices and videos, whereby as long as wireless equipments detect an event, the bursty traffic could be generated in short time scales. The self-similar traffic models characterize the statistical analysis results of data collected over a long time period, and thus the network traffic exhibits the properties of self-similarity [23]. But the self-similarity traffic models cannot be employed to predict the real-time network traffic for various applications. These two types of models can only characterize the packets with the same feature. However, in general networks scenarios, there are different network services while they usually have different kinds of packets such as the packets from the telemetry, the IoT, and the video streaming. Specially, the UAVs need to exhibit different network usage and behaviours to support various network services relying on packet arrival rate and packet size in UAV-enabled wireless networks. Accordingly, the network services should be divided into different subsets, and meanwhile the UAVs need to be categorized into different subgroups. Although these two issues are crucial to model the traffic flow, they have not been investigated in multi-services oriented UAV-enabled wireless networks by now.

To address the issues, this paper jointly considers these two issues in multi-services oriented UAV-enabled wireless networks. In particular, we calculate the number of packets and corresponding data size requested for each network service. The main contributions of this paper are summarized as follow.

- The network services relying on the telemetry, IoT and video streaming are categorized into three subsets, each of which has different packet transmission frequency and packet's size. We then use the Pareto distribution to model the non-uniformity feature of network usage among UAVs, whereby the UAVs are divided into three subgroups depending on their network usage. By combining these three subgroups and three subsets, nine network segments can be generated, each corresponding to a subgroup and a subset.
- We determine the average number of packets and the total data size in each segment, the number of transmitted packets and the data size in each network service.
- Finally, simulation results are provided to illustrate that our traffic model can well predict the number of packets and the data size for the different network services compared to these in real scenarios.

The rest of the paper is organized as follows. Section II introduce network services, subgroups for UAVs, the number of packets, and traffic data size. A case study of telemetry, IoT and streaming data scenario is discussed in section III. Simulation results are provided in section IV. Section V concludes this paper.

## II. Traffic Flow Modelling

This section models the traffic flow in multiple services oriented UAV-enabled wireless networks.

### A. Network Services for UAVs

The services provided by the network can be classified into three main subsets: telemetry data, IoT data and streaming data, which represent the information about the UAV's status and its component called the Telemetry data, the information packets from the IoT on-board the UAVs, and videos captured by the camera on-board the UAV, respectively

### B. Subgroups for UAVs

Depending on the frequency of the network usage, the UAV's behaviour can be classified into three subgroups according to the frequency from the lowest to the highest: 1) poor subgroup representing network users with very limited network usage, 2) middle subgroup representing average network users with a moderated network usage, and 3) rich subgroups representing network users that conduct a heavy usage on the network.

Pareto distribution [24] is commonly used to characterize the non-uniformity of income among population. It relies on the Gini coefficient [25] to calculate the number of participants in a given subgroup. We consider the Pareto distribution with parameter $\alpha_i$, where $i = 1, 2, 3$ represents the $i$th subset of network service.

Let $F_j$ denotes the proportion of UAVs in the $j$th subgroup, $j = 1, 2, 3$ represents the poor, the middle, and the rich subgroups, respectively. Lorenz curves corresponding to the Pareto distribution with parameter $\alpha$ can be written as follows.

$$Q(\alpha, x) = 1 - (a - F(x))^{\frac{\alpha-1}{\alpha}} \tag{1}$$

We assume that UAVs from the $3^{rd}$ subgroup will generate $90\%$ of video streaming data, and $90\%$ of IoT data will be generated by UAVs from the $2^{nd}$ and $3^{rd}$ subgroups. Then, using the Lorenz curves, the subgroups can be determined as [26]

$$F_3 = 0.9^{\frac{\alpha_3}{\alpha_3-1}}, \tag{2}$$

$$F_2 = 0.9^{\frac{\alpha_2}{\alpha_2-1}} - F_3, \tag{3}$$

and

$$F_1 = 1 - F_2 - F_3. \tag{4}$$

### C. Number of Packets

Two classes of events are formed to hold the previous two subgroups. The first class is for events from the three different network services. We use the index $i = 1, 2, 3$ to denote the telemetry, IoT, and stream data, respectively. The second class is for the three subgroups of network users. Denoted by $j = 1, 2, 3$ for the network users from poor, middle, and rich subgroups, respectively. The segregation of these two classes of events will generate nine segments, we calculate the packet arrival rate $\lambda_{ij}$ per UAV in each of them. Based on that, we further calculate the number of packets generated by a swarm of UAVs.

Let $\beta_{ij}$ denote the share of transactions requested for the $i$th subset of network services by users of the $j$th network subgroup. We define the share of transactions $\beta_{ij}$ in two different manners as demonstrated below.

*1) The share of transactions based on the frequency of requests for network services:* By combining the formulas (1), (2), (3), (4) and $\sum_j \beta_{ij} = 1$, we can calculate the values of $\beta_{ij}$ for the nine segments as follows.

$$\begin{aligned} \beta_{i1} &= 1 - (1 - F_1)^{\frac{\alpha_i-1}{\alpha_i}}, \\ \beta_{i2} &= 1 - (1 - F_1 - F_2)^{\frac{\alpha_i-1}{\alpha_i}} - \beta_{i1}, \\ \beta_{i3} &= 1 - \beta_{i1} - \beta_{i2}, \end{aligned} \tag{5}$$

*2) The share of transactions $\beta_{ij}$ based on the rate of transactions $\lambda_{ij}$:* Let $N$ be the total number of users in the network, and the total rate of transactions for the $i$th network service generated by all users of the $j$th subgroup can be defined as

$$S_{ij} = F_j N \lambda_{ij}, \tag{6}$$

where $S_i$ denotes the total rate of transactions.

The total rate of transactions generated by the requests for the $i$th network service can be defined as

$$S_i = N \sum_j \lambda_{ij} F_j \tag{7}$$

The expression of $\beta_{ij}$ can be found as the division of the two previous expressions as follow.

$$\beta_{ij} = \frac{S_{ij}}{S_i}. \tag{8}$$

Based on the initial input value of $\lambda_{11}$ which represents the average packet arrival rate for the telemetry data from a UAV of the poor subgroup, we can calculate the values of the following unknown parameters $\lambda_{ij}$.

$$\begin{aligned} \lambda_{12} &= \frac{\lambda_{11}\beta_{12}F_1}{\beta_{11}F_2}, \\ \lambda_{13} &= \frac{\lambda_{11}\beta_{13}F_1}{\beta_{11}F_3}, \\ \lambda_{ij} &= \frac{\gamma_i \beta_{ij} \sum_j \lambda_{1j}F_j}{\gamma_1 F_j}, i = 2,3; j = 1,2,3. \end{aligned} \tag{9}$$

Let $T$ denote the experiment duration in seconds. The total number of transmitted packets $P_i$ when requesting services from the $i$th network subset can be given as follows.

$$P_i = N \times T \sum_j \lambda_{ij}F_j. \tag{10}$$

### D. Traffic Data Size

We use $W_i$ to denote the average data size of a transaction from the $i$th network service subset. Then, the traffic data size from the transactions rate in each network subset can be determined as follows

$$D_i = N \times T \times W_i \sum_j \lambda_{ij}F_j, \tag{11}$$

where $D_i$ denotes the $i$th network subset.

We then calculate the total traffic data size as follows

$$D = N \times T \sum_i W_i \sum_j \lambda_{ij}F_j, \tag{12}$$

where $D$ denotes the total traffic data size.

## III. Case Study

In this section, we give two cases using our proposed network flow model, whereby the swarms of UAVs perform different tasks and request many network services aiming to capture and send real-time data to the end users.

### A. Weather Measurement and Video Streaming Scenario

In this scenario, these UAVs equipped with high definition cameras capture video streams along their path and send it to the end user in real-time. At the same time, these UAVs are also equipped with humidity and temperature sensors to gather weather measurements while following their mission path, and send them back to a remote server. Fig. 1 illustrates an example of the deployment of the UAVs. In Fig. 1, the red, green, and black UAVs represent that they are from the first, second, and third groups, respectively. The Wi-Fi sign represents the network usage in the position where the UAVs capture videos or send IoT data. The UAVs in the first group will not stream a video of their whole path but they will send the video in specific places. It is the same for the weather measurement. The UAVs in the second group correspond to network users, and they will stream video of some segments of their path and take weather measurement in many places. The UAVs in the third group conduct a heavy usage on the network by streaming a continuously streaming videos of their whole path and take weather measurement in all places.

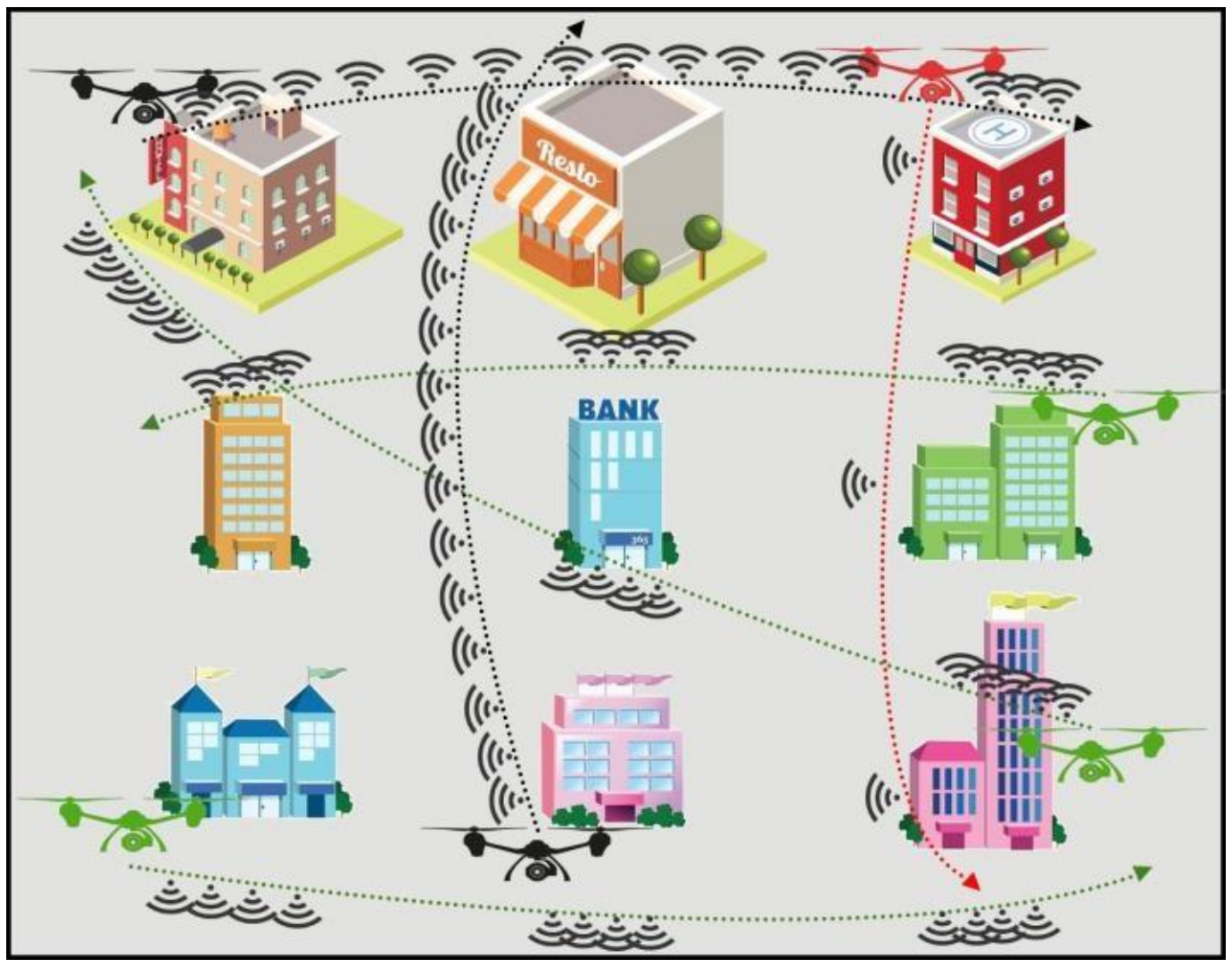

Fig. 1: Illustration of weather measurement and video streaming scenario

### B. BVLoS IoT Data-Collection Scenario

In this scenario, we use a swarm of UAVs to perform a specific BVLoS mission (Beyond-Visual-Line-of-Sight). Each UAV is equipped with a high-definition camera streaming a video to the remote pilot to be able to follow and control the UAV in real-time. These UAVs will be used to collect data from IoT devices located in some remote areas and send them back over the network to the end user. Since this mission is about the collection of the IoT data, the share of transaction for IoT packets will be approximately the double of the share of transactions in the previous scenario. Fig 2 shows an illustration of the deployment of the UAVs to perform the mission. The red, green, and black icons represents UAVs from the first, second, and third group, respectively. Fig 2 shows also the position of the IoT devices where the UAVs are supposed to go to collect data and send them over the network to the end user. The Wi-Fi sign represents the action of network usage by the UAV passing by that location. To optimize battery usage, UAVs from the first subgroup will stream videos only when they get to the location of the IoT devices. UAVs of the second subgroup will stream videos in some dangerous locations so the remote pilot can assist them to avoid obstacles. UAVs in the third subgroup will stream a video of the whole path of their mission, and the video streamed to the remote pilot can help him control the UAV's mission and plan its path.

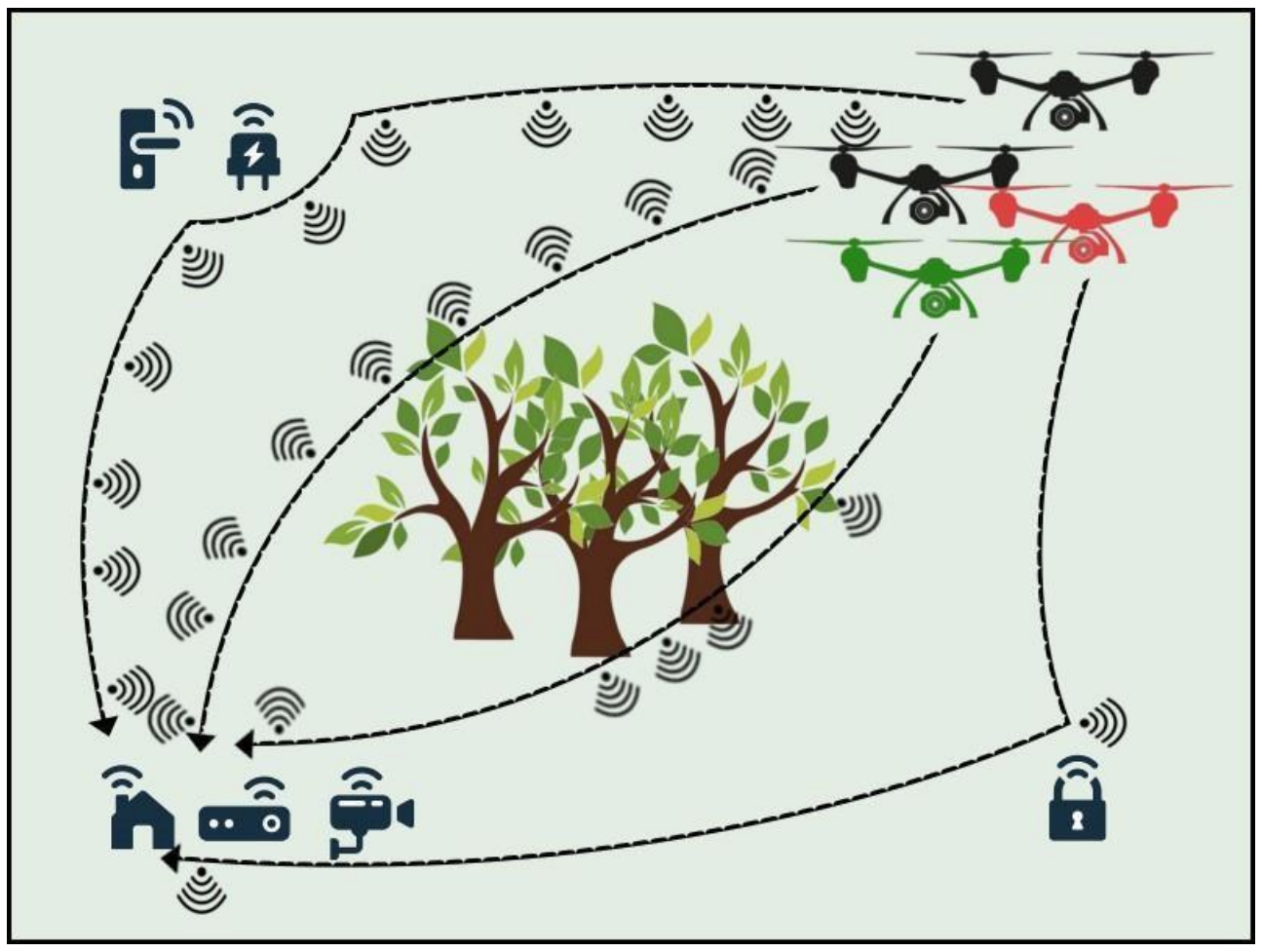

Fig. 2: Illustration of BVLoS IoT-data collection scenario

TABLE I: Parameter settings

| Parameters | Values |
|---|---|
| Pareto coefficient $\alpha_i$ for telemetry, IoT, and streaming data | 100, 1.125, 1.06 |
| Sharing ratio $\gamma_i$ for telemetry, IoT, and streaming data in the weather measurement scenario | 98%, 1.9%, 0.1% |
| Sharing ratio $\gamma_i$ for telemetry, IoT, and streaming data in the BVLoS IoT data-collection scenario | 96%, 3.9%, 0.1% |
| Average size $W_i$ of a transaction for telemetry, IoT, and streaming data | 96%, 3.9%, 0.1% |
| Total number of used UAVs | 20 |
| Observation duration in seconds | 60s |

## IV. Performance Evaluation

In this section, we will evaluate the performance of the scenarios illustrated previously in the case study.

### *A. Simulation Environment*

To simulate the case study presented in the section III, we use Software In The Loop (SITL) with Ardupilot to get the instances of UAVs. The tool can be used to simulate the behaviour of UAVs, interact with them, and listen for their telemetry data. We create 20 instance of UAVs as per the case study requirement, start these instances and collect their telemetry data that has been sent to our server using MAVLink protocol. IoT data for weather information is sent to our server using AMQP protocol. Video streams in the other side, are simulated using a Raspberry PI board that sends a high-definition videos to the server. We use only one unit to stream videos for all the UAVs. The unit behaves as only one video streamer at a time. We run the video streaming process as many time as much simulated UAVs we have in our scenario. In this simulation, 20 instances of UAVs have been created, each of them generates its own telemetry data and sends them to the remote server. With each UAV instance, a MAVLink packet listener is started in the specified port to collect the telemetry data. The script that creates the UAV instances, creates also a process of generating IoT data for this UAV and send them to the server using the AMQP protocol. An AMQP packet listener is started in the remote server to collect these packets. The process of generating video streams has been run separately, since we have 20 instances of UAVs, we run the streaming process on the Raspberry PI unit 20 times, each one corresponds to one UAV. The Raspberry PI unit sends the videos using ffmpeg tool to the server in which the videos are collected and the size is measured.

The parameters used in the simulation are summarized in Table I.

### *B. Performance Analysis of Weather Measurement and Video Streaming Scenario*

In this scenario, we analyze the number of telemetry, IoT, and video streaming packets transmitted over the network. We also calculate the total size of these packets predicted by our proposed theoretical traffic flow model. These packets are generated by a swarm of UAVs performing the data collection of weather information and sending video stream packets in a specific area. We further conduct a comparison study between theoretical and simulated results. Fig. 3 summarizes the total number of transmitted packets in theoretical and simulated cases. An observation from Fig. 3 shows that the number of telemetry packets is really higher than the number of IoT and streaming data. This is because the average telemetry packet rate is 100 packets per second per UAV while the average rate in the third subgroup (rich group) is about 21 and 0.6 packets per second per UAV for IoT and Streaming data, respectively. Note that these UAVs in the third subgroup uses the network more than the other groups. Fig. 4 shows that the streaming data represents the majority of the transmitted packets. This is due to the fact that the average data size of one video is extremely higher than the size of a telemetry and an IoT packet. Comparing Fig. 3 with Fig. 4, we can see that the theoretical results are nearly the same as the simulated results in the number and the size of all kind of network data which indicates that our traffic estimation model can predict the network traffic in real scenarios.

### *C. Performance Analysis of BVLoS IoT Data-Collection Scenario*

In this section, we analyze the behaviour of the UAVs in the network. This analysis comprises the calculation of the number of transmitted packet and their respective sizes in the given scenario. The analysed packets involve telemetry, IoT, and streaming data generated by the swarm of UAVs performing the BVLos IoT data-collection scenario. Then, we compare

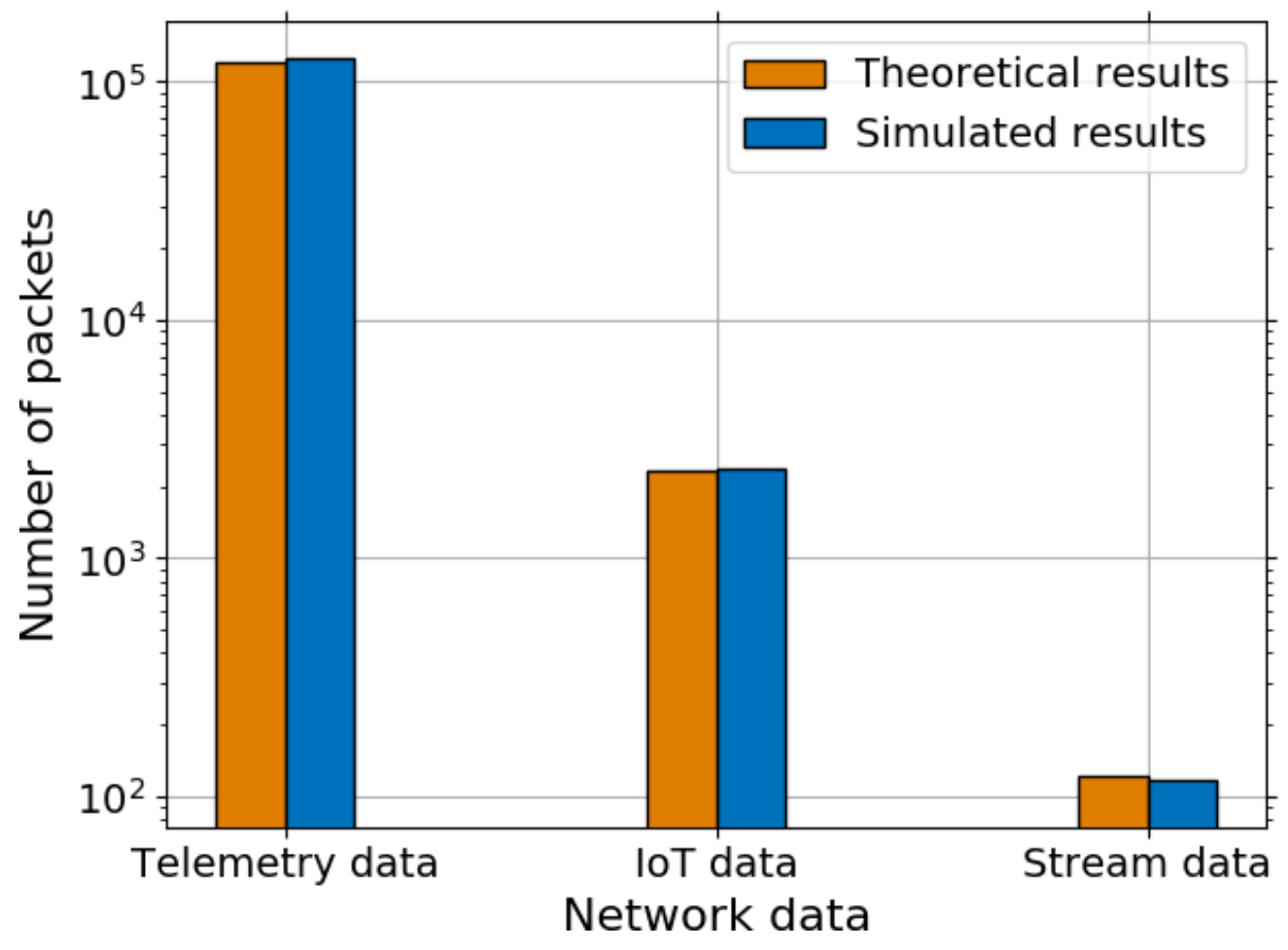

Fig. 3: Number of transmitted packets

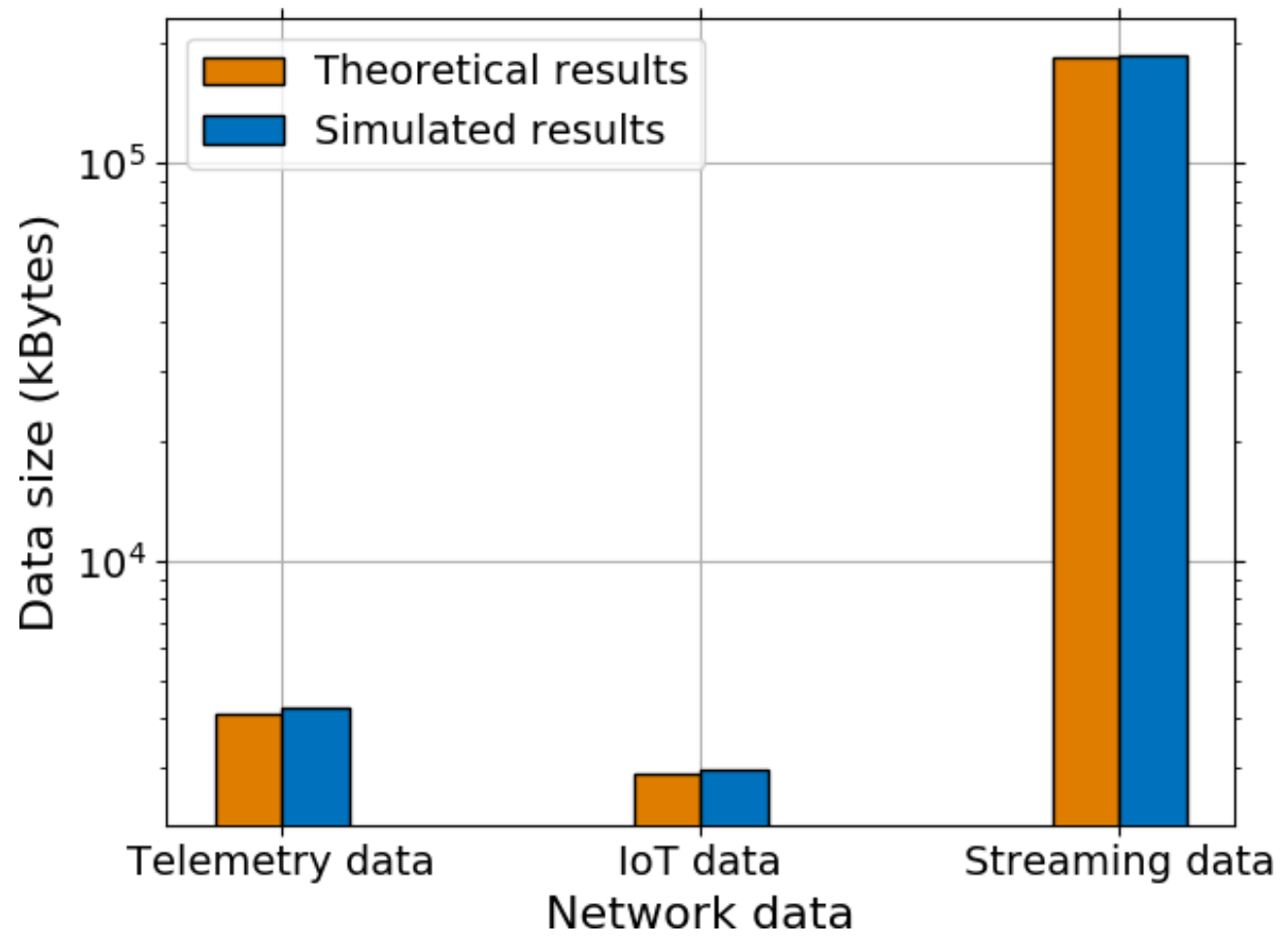

Fig. 4: Network data usage

these simulated data to the theoretical result we got using our model presented above.

We summarize in Fig. 5, the number of transmitted telemetry, IoT, and streaming packets in both theoretical and simulated cases. The figure is displayed in a log scale to better visualize the stream data values. Although, this scenario is for IoT data collection, the number of transmitted telemetry data exceeds the number of transmitted IoT data because the telemetry data transmission rate is 100 packets per second per UAV while the IoT data transmission rate is only 10 packets per second per UAV in the third user-subgroup.The number of transmitted videos is still very few compared to the other network data.

Fig. 6 illustrates the data size of the different transmitted network packets. We use a graph with a log scale to enlarge the telemetry data values for a better visualization. In this scenario, the size of transmitted IoT data exceeds the size of the telemetry data, this is because the number of transmitted IoT packets illustrated in Fig. 5 increases significantly compared to the previous scenario. The streaming data represents the majority of the network traffic, this is due to the fact that the average size of one video is extremely greater than the size of one telemetry or IoT packet.

Compared to the results in Fig. 5 and Fig. 6, we can see that the results from both the theoretical and the simulated scenarios are matched in number and in size of network packets for the telemetry, IoT, and video streams data.

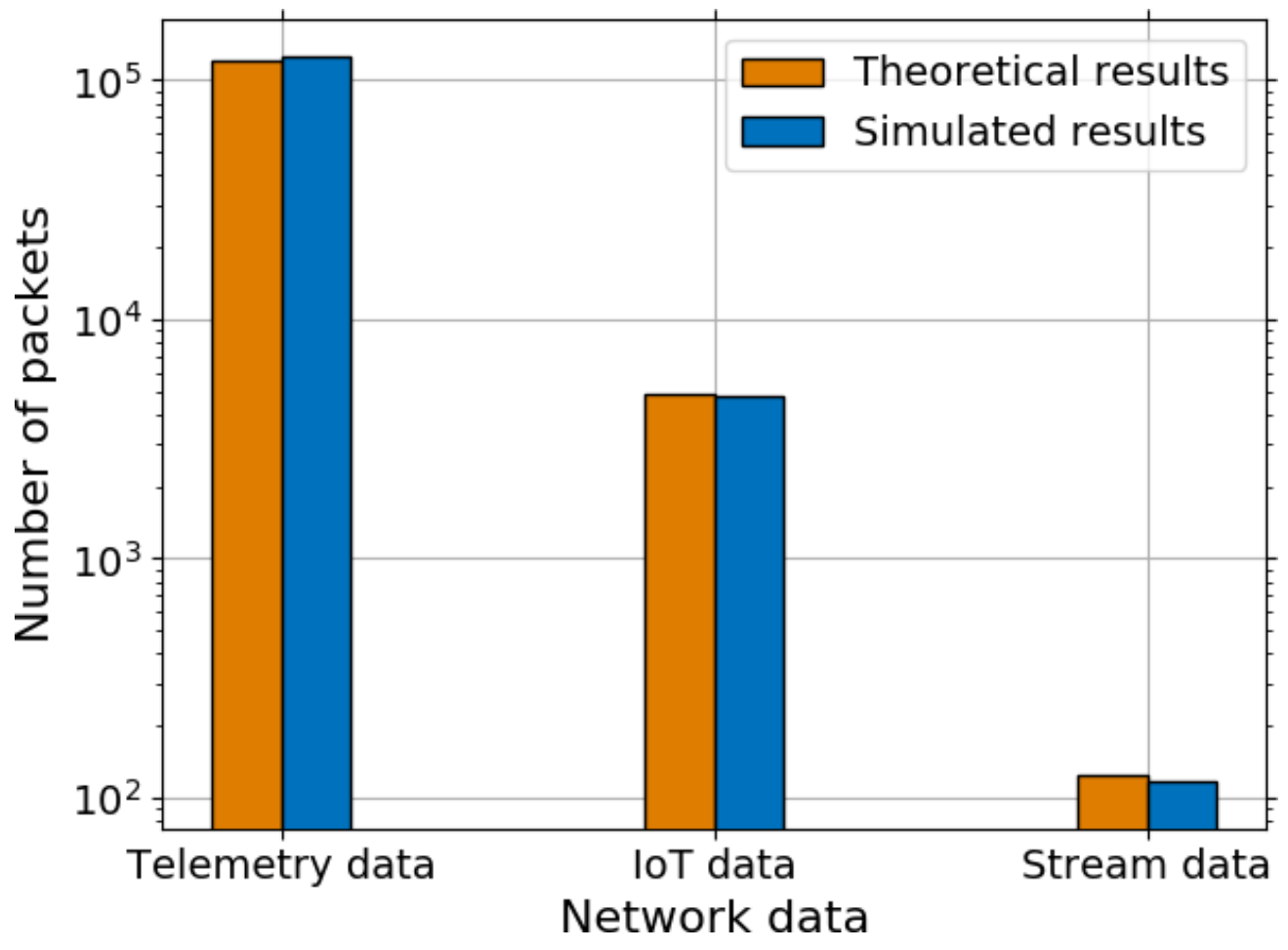

Fig. 5: Number of transmitted packets

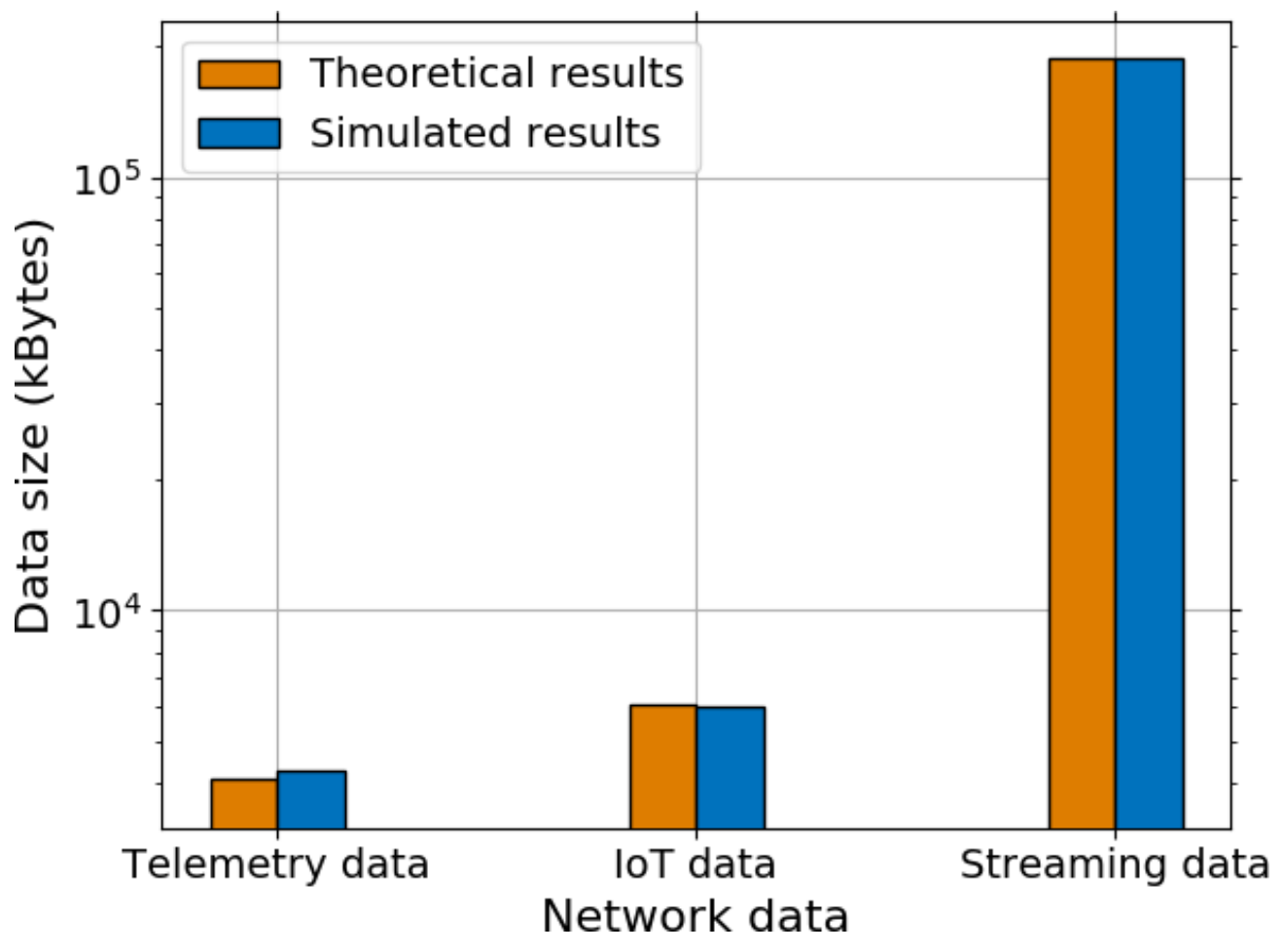

Fig. 6: Network data usage

## V. CONCLUSION

A general traffic flow model was proposed for the multiple services oriented UAV-enabled wireless networks. In this model, we first categorize the network services into three subsets, and then divide UAVs into three subgroups according to their network usage. We also provide the formulas for determining the number of packets for each service and total data size. The simulation results indicate that our proposed model is suitable for predicting the traffic data in the network for the different monitored network services.

## ACKNOWLEDGEMENT

This work was supported by the European Union's Horizon 2020 Research and Innovation Program through the 5G!Drones Project under Grant No. 857031, the Academy of Finland 6Genesis project under Grant No.318927, the Academy of Finland CSN project under Grant No. 311654, the NSF of China under Grant No. 61962033, the Anhui Province project under Grant No. 1808085MF165, gxgwfx2019060 and KJ2019A0643, and the Yunnan Province project under Grant No. 2018FH001-010.